\documentclass[12pt]{iopart}
\usepackage{graphicx}
\newcommand{ \srt  }{$\sqrt{s_{_{\rm NN}}}$}
\newcommand{ \mpt } {$\langle p_T \rangle$}
\def\Journal#1#2#3#4{{#1} {\bf #2}, #3 (#4)}
 
\def\NPA{{\em Nucl. Phys.} A}
\def\PLB{{\em Phys. Lett.}  B}
\def\PRL{\em Phys. Rev. Lett.}
\def\PRC{{\em Phys. Rev.} C}

\def\JPG{{\em J. Phys.} G}
\def\EPJ{{\em Eur. Phys. J.} C}

\begin{document} 
\title{Bulk properties and flow}
\author{Zhangbu Xu} 
\address{Physics Department, Brookhaven National
Laboratory, Upton, NY 11973}

\date{\today}
\begin{abstract}
In this report, I summarize the experimental results on {\bf bulk
properties and flow} presented at Quark Matter 2004. It is organized
in four sections: 1) Initial condition and stopping; 2) Particle
spectra and freeze-outs; 3) Anisotropic flow; 4) Outlook for future
measurements.
\end{abstract}
A wealthy set of data related to bulk properties and flow had been
presented at this conference. I will highlight on the results related
to initial condition, partonic collectivity and thermalization in
relativistic heavy ion collisions, emphasizing on the exciting results
from recent RHIC run with Au+Au and d+Au collisions. The plots shown
in this report are to illustrate the discussions and I refer the
readers to presentations by various collaborations for complete
dataset. Other important and exciting results, such as strangeness
($K/\pi$) excitation function from SPS~\cite{NA49} and pentaquark
searches from various experiments~\cite{pentaquark}, will not be
covered in this summary.
\section{Initial conditions}
One of the exciting results in forward rapidity (deuteron side) at
RHIC is the observation of suppression of particle yields at and below
intermediate $p_T$~\cite{brahmscgc} in d+Au collisions. The
observations~\cite{brahmscgc,phenixcgc,phoboscgc,starcgc} are
consistent with the qualitative derivation from the theory of Color
Glass Condensate ({\bf CGC}). Lively discussions had been ongoing
about whether models with conventional stopping and/or shadowing can
explain the effects~\cite{guylassy,nagle}. Conclusive evidences may
have to wait for more quantitative predictions from theory and
experimental observations of more universal scalings related to the
unique physical quantity of saturation momentum scale ($Q_s$). PHOBOS
Collaboration has measured the charged particle multiplicity
distributions in the full pseudo-rapidity ($-5<\eta<5$) range in p+p
collisions and several centralities of the d+Au and Au+Au
collisions~\cite{phobos,roland}.  BRAHMS, PHENIX and STAR
collaborations have presented results in various $\eta$ coverage as
well. The results appear to be consistent with HIJING/AMPT models,
limiting fragmentation and in disagreement with an earlier prediction
by saturation models~\cite{phobos,brahmscgc}. Updated predictions from
saturation model agree better with the experimental results in the
d+Au centrality dependence~\cite{dimadAu}. BRAHMS presented the identified
particle yield as function of rapidity and transverse
momentum~\cite{murray}. From the produced particle yields and net
baryon rapidity loss, it was concluded that projectile Au nuclei lose
$25\pm1$ TeV in the central Au+Au collisions. This is about 75\% of
the beam energy. The rapidity loss is also a relevant factor in the
interpretation of the suppression in the forward rapidity. The $\pi$
rapidity distribution is Gaussian with width of $\sigma=2.26\pm0.02$
and consistent with simple estimate from Landau hydrodynamics.  The
anisotropic flow $v_2$ as function of $\eta$ measured by PHOBOS
Collaboration also shows that it is quite Gaussian.  These are very
convincing evidences that the rapidity distribution of the particles
($\rightarrow$energy) and its subsequent expansion is NOT Bjorken
boost invariance. There was debate at the conference about the
validation of the models with Bjorken boost invariance at the limit 
rapidity range (i.e. $|y|<1$).

\section{Particle spectra, freeze-outs, and nuclear modification factor}
The statistical model has been able to describe the yield ratios of
the identified particles successfully~\cite{pbm}. The question is what
it means since the same model can fit well to those in p+p and AA
collisions~\cite{pbm}. To distinguish a thermal system from a system
born with the thermal phase space where temperature is simply a
Lagrange multiplier~\cite{koch}, we need to measure the interactions
between particles and observe the evolution of the system from
canonical to grand canonical ensemble. The interaction of particles
was measured through resonance yields according to their hadronic
re-scattering cross section and collective flow of many identified
particles.  The evolution of the system from canonical to grand
canonical ensemble was observed by the centrality dependence of the
canonical suppression of strangeness in small systems.
\begin{figure}[ht]
\begin{center}
\includegraphics*[keepaspectratio,scale=0.55]{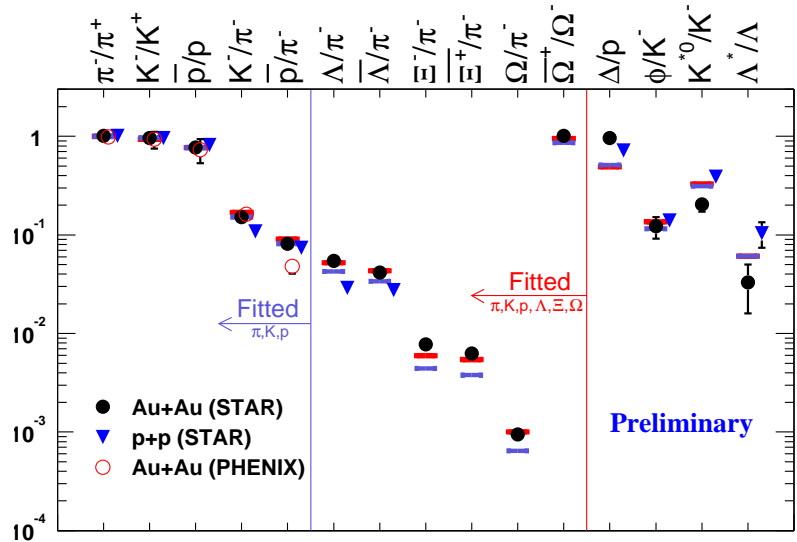}
\caption{Particle ratios in p+p and central Au+Au collisions at
RHIC. The filled circles (open circles) are ratios in central Au+Au
collisions measured by STAR (PHENIX) Collaboration.  The triangles are
particle ratios in p+p minbias collisions measured by STAR
Collaboration.  The red (light blue) bars are results from statistical
model fit to $\pi,K,p,\Lambda,\phi,\Xi$ and $\Omega$ ($\pi,K,p$). }
\label{Fig:ratios}
\end{center}
\end{figure}
\subsection{Particle yield ratios and chemical freeze-out}

Both SPS and RHIC have measured a rich spectrum of particle including
$\pi,K,p,\Lambda,\phi,\Xi,\Omega$ in p+p and many centralities in AA
collisions~\cite{antinori}. Fig.~\ref{Fig:ratios}(left portion) shows
the variety of the stable particle ratios in p+p and central Au+Au
collisions at \srt $=200$ GeV. The filled (open) circles are
measurements by STAR (PHENIX) in Au+Au collisions. The ratios measured
by PHOBOS and BRAHMS~\cite{phobosratios,brahmsratios} are consistent
with the measurements plotted in Fig.~\ref{Fig:ratios}. The filled
triangles depict STAR's results from p+p collisions. The red (light
blue) bars correspond to statistical model fit to
$\pi,K,p,\Lambda,\phi,\Xi$ and $\Omega$ ($\pi,K,p$) in Au+Au
collisions~\cite{olga,kai}. I list a few observations:
\begin{itemize}
\item The statistical model~\cite{olga} fit to
   $\pi,K,p,\Lambda,\phi,\Xi$ and $\Omega$ results in
   $T_{ch}=160\pm5$, $\mu_B = 24\pm4$, $\mu_S=1.4\pm1.6$ and
   strangeness saturation factor $\gamma_S = 0.99\pm0.07$.
\item Strange particles are enhanced in Au+Au over p+p
      collisions~\cite{pbm}. This is also evident from the strangeness
      enhancement factor as function of centrality~\cite{antinori}.
\item The fit to $\pi, K, p$ can not constrain the strangeness
parameters to predict well the multistrangeness particle production.
\item The $\bar{p}/\pi$ ratio from PHENIX is 40\% lower than that from
      STAR due to different weak-decay feed-down correction. STAR
      measures inclusive $\bar{p}$ production with PDG
      definition~\cite{starpbar,PDG} (P226 ``the rates given include
      decay products from resonances with $c\tau<10$ cm'').  PHENIX's
      measurements are weak-decay feed-down corrected with $\Lambda$
      from measurements and $\bar{\Sigma}^{-}$ from
      model~\cite{phenixpid}.
\end{itemize}
The centrality dependence of ${S/(Npart/2)}_{AA}/S_{pp}$ where $S$ is
either $\Lambda,\Xi$ or $\Omega$ increases monotonically and shows no
sign of saturation.  However, the charged particle density which is
more propotional to the volume of the system doesn't scale linearly
with $Npart$ from p+p to central Au+Au collisions. It was noticed that
the $\Lambda$ yield scales linearly with the charged particle density
in Au+Au collisions~\cite{starlambda}.  The discussion at the
conference was centered at the questions of whether the number of
participants $Npart$ is the right reference as a volume indicator and
whether the system reaches grand canonical ensemble. The fact that the
strangeness saturation factor is consistent with
unity~\cite{pbm,olga,kai} suggests that the grand canonical ensemble
is reached at least in the central collisions.
\subsection{Resonances as a probe of freeze-out dynamics}
Also shown in Fig.~\ref{Fig:ratios} are the ratios of resonance yields
over stable particle yields. The resonances are $\phi, K^*, \Delta,
\Lambda^*$ with ratios of $\phi/K,K^*/K,\Delta/p$, and
$\Lambda^*/\Lambda$. These ratios were also shown to have different
centrality dependence~\cite{haibin,markert,fachini}. When compared to
statistical model predictions from fitting results of stable
particles, the ratios of $K^*/K$ and $\Lambda^*/\Lambda$ in central
Au+Au collisions are low. At the same time, the $\phi/K$ ratio is very
close to the prediction and independent of centrality while
$\Delta^{++}/p$ ratio is a little higher than but within the error of
the prediction. We reached same conclusion by comparing the results
from Au+Au to those from p+p collisions. This indicates that the
peculiar feature is model-independent.  A natural explanation to these
behaviors will be the re-scattering and regeneration of resonances via
hadronic interactions. The system in the hadronic phase is effectively
a pion gas. Resonances with short lifetime ($c\tau\simeq$ fm) will
decay in an environment surrounding by pions. The daughters from these
decays will re-scatter with pions and other particles. On the other
hand, the hadrons will interact to generate the resonances through the
reverse reaction channel of the decay. In the case of $K^*$, the
reaction channel will be $K^*\longleftrightarrow\pi K$ and the
re-scattering channel is $\pi\pi\rightarrow\rho$. Since the $\pi\pi$
cross section is large ($\sigma_{\pi\pi}\simeq5 \sigma_{\pi K}$), it
is a good assumption that the daughter pions from $K^*$ decay before
kinetic freeze-out will not escape the system.  Therefore, we will not
be able to observe those $K^*$ decay in the hadron gas before kinetic
freeze-out.  The reaction channels of $\phi\leftrightarrow KK$,
$K^*\leftrightarrow K\pi$,$\Delta\leftrightarrow p\pi$, and
$\Lambda^*\leftrightarrow K\Lambda$ and their daughter rescatterings
have very different cross seection and the yield ratios of
$\phi/K,K^*/K,\Delta/p$, and $\Lambda^*/\Lambda$ are consistent with
the cross sections. We conclude:
\begin{itemize}
\item Interactions between particles happen in the system.
Temperature from the thermal fit is unlikely to be just a lagrange
multiplier~\cite{koch}.
\item The balances of resonance re-scattering and re-generation are
not maintained in the course of the expansion. The resonance yields
reflect the time span of the expansion~\cite{haibin,markert,fachini}.
\item Interactions happen with finite cross section comparable to the
cross section measured and implemented in transport models~\cite{fachini,bleicher}.
\item It is too idealized to assume infinite cross section in
hydrodynamics all the way to the kinetic freeze-out.
\item It is too idealized to assume that the same chemical equilibrium
in the resonance yields at the chemical freeze-out can be maintained
up to the kinetic freeze-out.
\end{itemize}

\begin{figure}[ht]
\begin{center}
\includegraphics*[keepaspectratio,scale=0.55]{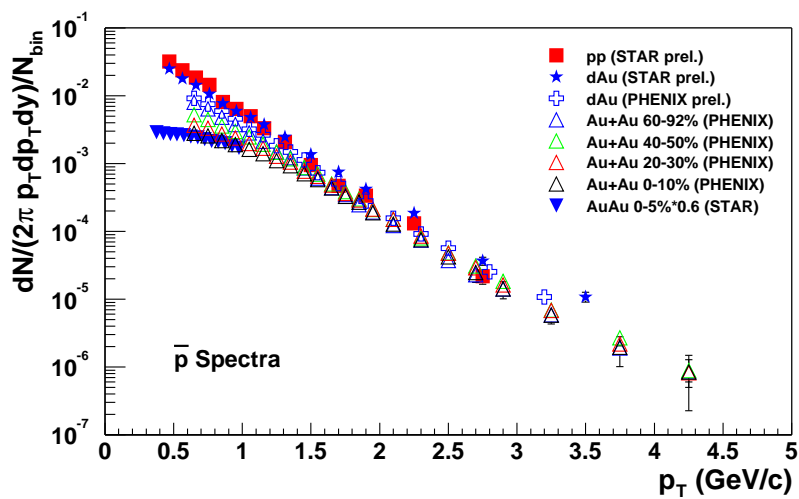}
\caption{Anti-proton spectra from p+p, dAu and Au+Au collisions at
\srt 200 GeV.  Squares and stars are results from p+p and d+Au,
respectively. The triangles are different centrality bins in Au+Au
collisions.}
\label{Fig:pbar}
\end{center}
\end{figure}
\subsection{Spectra, radial flow, nuclear modification factor, and Cronin Effect}
The identified particle spectra have been used to extract information
about radial flow and kinetic freeze-out. The main focus is on the
issue of whether there is a single freeze-out and if not, are the
particles with different hadronic scattering cross sections decoupling
at different time? Detailed discussions have been
presented~\cite{olga,murray,fabrice,lisa} with comparisons/fits of
models of hydrodynamics or hydro-inspired blast wave to the measured
identified particle spectra. Although some of the results indicated
that hydrodynamics with one single freeze-out describes the spectra
well, the others showed that sequential freeze-out of different
particles is needed to fit the data. The discussions raised to the
level of whether we are ``abusing'' (relying too much on,
over-interpreting) the model comparisons~\cite{lisa}. However, the
sequential freeze-out is more consistent with the rescattering of
resonance discussed in previous section. Fig.~\ref{Fig:pbar} showed
the $\bar{p}$ spectra from p+p, d+Au and Au+Au collisions by the four
RHIC experiments~\cite{ruan,matathias,veres}.  The plot shows that
from p+p to all centralities in Au+Au collisions, the $\bar{p}$
spectra exhibit binary scaling at $p_T{}^>_{\sim}2$ GeV/c. The
centrality difference is prominent at low $p_T$ where the spectrum is
flatter from p+p to central Au+Au collisions. BRAHMS showed that there
is very little centrality dependence in the spectral
shape~\cite{murray}, therefore centrality dependence of radial flow is
not as dramatic. However, this depends on the $p_T$ and centrality
coverage.  The feature of low yield at low $p_T$ and binary scaling at
high $p_T$ is counter-intuitive to the radial flow effect since one
would have expected the yields at high $p_T$ to increase from
peripheral to central due to the flow push of low $p_T$ protons to
higher $p_T$. This also shows that the spectrum at $p_T{}^>_{\sim}2$
GeV/c is not sensitive to the flow effect and probably should not be
included in the simple blast wave fit or hydrodynamic model comparison.

Suppression of high $p_T$ hadron production has been observed at RHIC
in central Au+Au collisions relative to p+p collisions. This
suppression has been interpreted as energy loss of the energetic
partons traversing the produced hot and dense meduim. At intermediate
$p_T$, the degree of suppression depends on particle
species. Surprisingly, at the range of $2{}^<_{\sim} p_T {}^<_{\sim}5$
GeV/c the baryons appear to scale approximately with number of binary
collisions in Au+Au collisions relative to p+p collisions while the
mesons are greatly suppressed.  The new data presented in this
conference tried to address the following questions:
\begin{itemize} 
\item whether the effect is a baryon-meson or a mass effect. This was
achieved by comparing the nuclear modification factors of baryons
($p,\Lambda,\Xi,\Omega$), light mesons ($\pi,K,\eta$) to heavy mesons
($\phi, K^*$).
\item whether the effect is a final state or initial state
effect. This was achieved by comparing the nuclear modification
factors of different particle species in Au+Au to d+Au collisions.
\item what hadronization scheme can explain the experimental
observations in the most ecomonic way.
\end{itemize}
\begin{figure}[ht]
\begin{center}
\includegraphics*[keepaspectratio,scale=0.55]{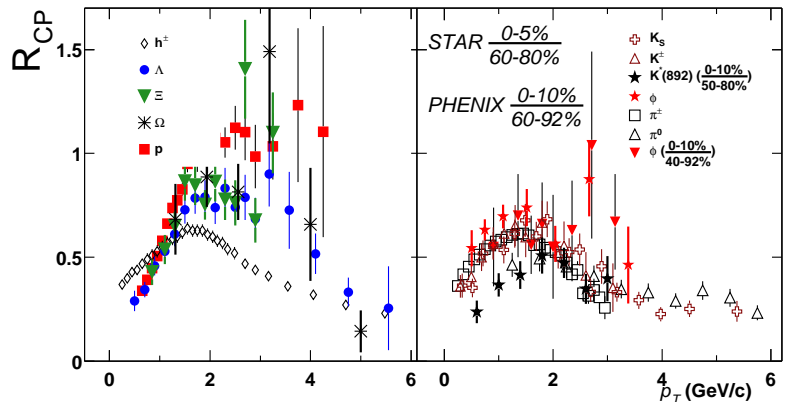}
\caption{$R_{CP}$ (central/peripheral) of
$\pi,K,p,K^*,\phi,\Lambda,\Xi$ and $\Omega$.}
\label{Fig:rcp}
\end{center}
\end{figure}

Fig.~\ref{Fig:rcp} shows the $R_{CP}$ (central/peripheral) of
$\pi,K,p,K^*,\phi,\Lambda,\Xi$ and $\Omega$ presented by STAR and
PHENIX collaborations. Although higher statistics are necesary for
particles, such as $\phi$ and $\Omega$, it is quite convincing that
the $R_{CP}$ exhibits a baryon-meson effect from the comparison. One
of the highlights in this conference is the successful explanation of
$v_2$ scaling and this intriguing result from
coalescence/recombination at the quark level~\cite{fries}.  Nuclear
modification factors in d+Au
($R_{dAu}$)~\cite{ruan,matathias,startof,barnby} have been presented
in the conference and publications as well~\cite{startof}. It showed
that the proton ($\Lambda$) $R_{dAu}$ at the intermediate $p_T$ is
larger than unity as also evident from Fig.~\ref{Fig:pbar}. This
contrasts to the binary scaling in Au+Au collisions. The difference
between $R_{dAu}$ of pions and protons is about 20\% while $R_{AuAu}$
are different by a factor of $\sim4$~\cite{startof,matathias}.  It was
concluded that the particle dependence of the suppression observed in
Au+Au collisions is also a final state effect. The good candidate
mechanism is the coalescence/recombination at the quark
level~\cite{fries,hwa}.
\section{Anisotropic flows}
\begin{figure}[ht]
\begin{center}
\includegraphics*[keepaspectratio,scale=0.45]{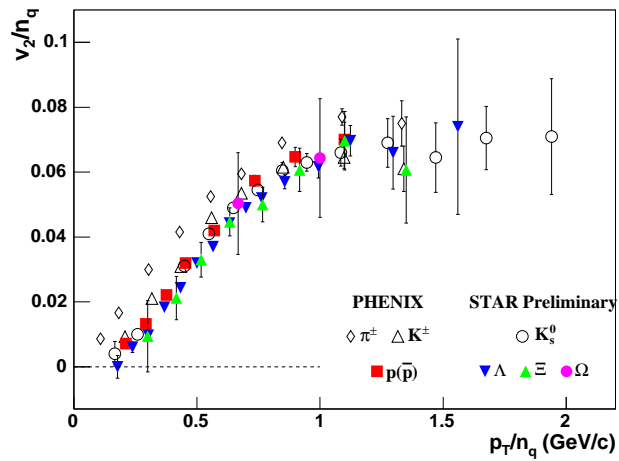}
\caption{$v_2$ of identified particles as function of $p_T$. Both axes
are scaled by the number of constituent quarks.}
\label{Fig:v2}
\end{center}
\end{figure}

The anisotropic flows measure the asymmetry of particle density in
momentum space relative to the reaction plane. The geometry is
initially asymmetric in non-central AA collisions.  The transformation
from geometric asymmetry to momentum-space asymmetry requires strong
interactions at early stage where the geometry has not been blurred by
the expansion. Anisotropic flow is one of the key achievements of the
RHIC bulk program where good agreement between measured $v_2$ and
ideal hydrodynamical calculations was reached~\cite{shuryak0312}.  The
conference witness beautiful set of data including:
\begin{itemize}
\item Higher harmonics: $v_1, v_2, v_4$~\cite{aihong,art,starv1,phobosv1}.
\item Different particle species: $e,\gamma,\pi,
K,p,\Lambda,K^*,\phi,\Xi,\Omega$~\cite{haibin,aihong,kaneta,paul,javier}. 
\item Large coverage in rapidity and centrality:
$-5<\eta<5$~\cite{aihong,phobosv1}.
\item Non-flow, cumulant, and newer techniques~\cite{aihong,markus}. 
\end{itemize}
We saw a set of $v_2$ measurements with not only $\pi,
K,p,\Lambda,K^*,\phi$ but also multistrange particles $\Xi$ and
$\Omega$. More importantly, the measured $v_2$ values as function of
$p_T$ scale with the number of constituent quarks as shown in
Fig.~\ref{Fig:v2}. This indicates that the number of the constituent
quarks in a hadron is a relevant degree of freedom as in the case of
nuclear modification function~\cite{fries,hwa}.  In addition, hadrons
with strange quarks behave the same as other particles.  It has been
argued that $\Omega$ flow will be a signature of partonic
collectivity~\cite{derekhydro,xu2}. Now $\Omega$ and $\Xi$ have large
$v_2$ which comes from early stage of the interaction. At the hadron
phase, hadrons interact according to the hadron-hadron cross section
as observed from resonances. Keep in mind that all the $\Omega$
excited states do not decay back to $\Omega$. Therefore, any violent
interactions need to drive the $\Omega$ flow at hadron stage will
result in disappearance of $\Omega$.  The only natural and reasonable
explanation for the $\Omega$ $v_2$ is that it is from partonic stage.

\section{Outlook for future measurements}

\begin{enumerate}
\item Forward rapidity. \\ Looking forward is always a good idea!
Results of forward spectra, jet and heavy flavor will be important to
study the effect of gluon saturation.
\item High statistics: $\phi,\eta,K^*,\Omega$ $R_{AA}$ and $v_2$. \\
Run 4 at RHIC will enable us to have high statistic sample to study 
$R_{AA}$ and $v_2$. 
\item Charm flow as a probe of thermalization.\\ PHENIX showed
promising results from $v_2$ measurements of charm via its
semileptonic decay~\cite{kaneta}. With direct open charm and its
semileptonic decay, we would expect exciting results of radial flow
and $v_2$ results from RHIC run 4 by STAR and PHENIX collaborations at
QM05.
\item Identified particle $v_1,v_2,v_4$. \\ STAR showed that there are
simple scalings between different harmonics~\cite{art,starv1}.  Look
forward to seeing the results from identified particle $v_1,v_4$ at
QM05.
\item Identified particle spectra in the away-side jet. How can we use
the jet quenching to study the thermalization of the system? \\ One of
the highlights at this conference is the experimental capability of
reconstructing the soft particle spectra from a jet traversing a
medium in Au+Au collisions at RHIC~\cite{fqwang}.  The spectral
multiplicities and total scalar $p_T$ in both near-side and away-side
increase with centrality (toward central) accompanying the
disappearance of back-to-back correlation at higher $p_T$ of the
associated particles. Moreover, it was shown that the away-side
spectrum becomes softer from p+p toward central Au+Au collisions while
the bulk spectrum becomes harder. We can roughly estimate the
thermalized fraction of away-side jet from the \mpt \; at central
Au+Au collisions by \mpt$_{AuAu, jet} = 80\%$\mpt$_{AuAu,
bulk}+20\%$\mpt$_{pp, jet}$. Similar to what discussed in previous
section, if the chemical composition (e.g. strangeness content) of the
away-side soft spectra changes from canonical ensemble in p+p
collisions to grand canonical ensemble in central Au+Au collisions,
this will be a strong signature that the jet probe ``melts'' and
becomes part of the thermalized system. One should look forward to
this at QM05.
\item Forward-backward rapidity asymmetry in d+Au.\\ It is still an
ongoing research on whether the Cronin Effect seen in d+Au collisions
can be explained by final state effect as well~\cite{hwa}. Further
experimental results on particle dependence and forward-backward
rapidity asymmetry will help resolve the
issue~\cite{hwa,wang,accardi}.
\item Energy Scan.\\ It will be very important/interesting to see
whether some of the key phenomena have any energy dependence.
\end{enumerate}

\section{Conclusions}
Intriguing physics in forward rapidity in d+Au collisions has been
observed at RHIC. Thermal/statistical model describes well the stable
particle yields up to $\Omega$ at mid-rapidity in central Au+Au
collisions. The radial flow derived from the identified particle
spectra and the finite hadronic scattering cross section from the
resonance yields indicate that the chemical freeze-out is followed by
a kinetic freeze-out.  It is clear that the measurements of nuclear
modification function and anisotropic flow ($v_2$) of the identified
particles support parton coalescence/recombination for particle
production at the intermediate $p_T$. The strong anisotropic flows of
identified particles up to $\Omega$ together with the observation of
finite hadronic cross section from resonances are a strong indication
of partonic collectivity. In the near future, the measurements will 
emphasize on thermalization and detailed evolution of the system. 

\ack The author would like to thank the organizers for the invitation
and thank L. Barnby, M. Bleicher, J. Castillo, X. Dong, P. Fachini,
J. Gonzalez, T. Hallman, H.Z. Huang, D. Kharzeev, M. Kaneta,
F. Matathias, L. McLarren, M. Murray, M. Oldenburg, J. Rafelski,
F. Retiere, L.J. Ruan, K. Schweda, J. Sandweiss, P. Sorensen, A. Tang,
T. Ullrich, G. Veres, F.Q. Wang, X.N. Wang, N. Xu, A. Tai and
H. Zhang.

\end{document}